\documentclass[prl,twocolumn,showpacs,floatfix]{revtex4}
\usepackage{graphicx}%
\usepackage{dcolumn,float}

\begin{document}

\title{Russian Doll Renormalization Group and Superconductivity}
\author{Andr\'e  LeClair$^\dagger$,
Jos\'e  Mar\'{\i}a Rom\'an and Germ\'an Sierra}
\affiliation{Instituto de F\'{\i}sica Te\'orica, UAM/CSIC, Madrid, Spain.} 
\date{November, 2002}

\begin{abstract}
We show that an extension of the standard BCS
Hamiltonian leads to an infinite
number of condensates with different energy gaps 
and self-similar properties,
described by a cyclic RG flow of the BCS coupling constant
which returns to its original value after a finite RG time. 
\end{abstract}

\pacs{11.10.Hi, 74.20.Fg, 71.10.Li}

\maketitle

\vskip 0.2cm

%		DEFINITIONS FOR TEX
%
%%%%%%%%%%%%%%%%%%%%%%%%%%%%%%%%%%%%%%%%%%%%%%%%%%%%%%%%%%%%%%%
%
%
%\def\e{\'e}
%\def\ee{\`e}
%%%%%%%%%%%%%%%%%%%DEFINITIONS%%%%%%%%%%%%%%%%%%%%%%%%%%%%%%%%%
%
\def\oti{{\otimes}}
\def\lb{ \left[ }
\def\rb{ \right]  }
\def\tilde{\widetilde}
\def\bar{\overline}
\def\hat{\widehat}
\def\*{\star}
\def\[{\left[}
\def\]{\right]}
\def\({\left(}		\def\BL{\Bigr(}
\def\){\right)}		\def\BR{\Bigr)}
	\def\BBL{\lb}
	\def\BBR{\rb}
%
%%%%%%%%%%%%%%%%%%%%%%%%%%%%%%%%%%%%%%%%%%%%%%%%%%%%%%%%%%%%%%%
%
\def\zb{{\bar{z} }}
\def\zbar{{\bar{z} }}
\def\frac#1#2{{#1 \over #2}}
\def\inv#1{{1 \over #1}}
\def\half{{1 \over 2}}
\def\d{\partial}
\def\der#1{{\partial \over \partial #1}}
\def\dd#1#2{{\partial #1 \over \partial #2}}
\def\vev#1{\langle #1 \rangle}
\def\ket#1{ | #1 \rangle}
\def\rvac{\hbox{$\vert 0\rangle$}}
\def\lvac{\hbox{$\langle 0 \vert $}}
\def\2pi{\hbox{$2\pi i$}}
\def\e#1{{\rm e}^{^{\textstyle #1}}}
\def\grad#1{\,\nabla\!_{{#1}}\,}
\def\dsl{\raise.15ex\hbox{/}\kern-.57em\partial}
\def\Dsl{\,\raise.15ex\hbox{/}\mkern-.13.5mu D}
%
%%%%%%%%%%%%%%%%%%%%GREEK LETTERS%%%%%%%%%%%%%%%%%%%%%%%%%%%%%%
%
\def\th{\theta}		\def\Th{\Theta}
\def\ga{\gamma}		\def\Ga{\Gamma}
\def\be{\beta}
\def\al{\alpha}
\def\ep{\epsilon}
\def\vep{\varepsilon}
\def\la{\lambda}	\def\La{\Lambda}
\def\de{\delta}		\def\De{\Delta}
\def\om{\omega}		\def\Om{\Omega}
\def\sig{\sigma}	\def\Sig{\Sigma}
\def\vphi{\varphi}
%
%%%%%%%%%%%%%%%%%%%CALIGRAPHIC LETTERS%%%%%%%%%%%%%%%%%%%%%%%%%
%
\def\CA{{\cal A}}	\def\CB{{\cal B}}	\def\CC{{\cal C}}
\def\CD{{\cal D}}	\def\CE{{\cal E}}	\def\CF{{\cal F}}
\def\CG{{\cal G}}	\def\CH{{\cal H}}	\def\CI{{\cal J}}
\def\CJ{{\cal J}}	\def\CK{{\cal K}}	\def\CL{{\cal L}}
\def\CM{{\cal M}}	\def\CN{{\cal N}}	\def\CO{{\cal O}}
\def\CP{{\cal P}}	\def\CQ{{\cal Q}}	\def\CR{{\cal R}}
\def\CS{{\cal S}}	\def\CT{{\cal T}}	\def\CU{{\cal U}}
\def\CV{{\cal V}}	\def\CW{{\cal W}}	\def\CX{{\cal X}}
\def\CY{{\cal Y}}	\def\CZ{{\cal Z}}

\def\rvac{\hbox{$\vert 0\rangle$}}
\def\lvac{\hbox{$\langle 0 \vert $}}
\def\comm#1#2{ \BBL\ #1\ ,\ #2 \BBR }
\def\2pi{\hbox{$2\pi i$}}
\def\e#1{{\rm e}^{^{\textstyle #1}}}
\def\grad#1{\,\nabla\!_{{#1}}\,}
\def\dsl{\raise.15ex\hbox{/}\kern-.57em\partial}
\def\Dsl{\,\raise.15ex\hbox{/}\mkern-.13.5mu D}
%
%%%%%%%%%%%%%%%%%%%%GREEK LETTERS%%%%%%%%%%%%%%%%%%%%%%%%%%%%%%
%
%%%%%%%%%%%%%%% MATH CHARACTERS %%%%%%%%%%%%%%%%%%%%%%%%%%%%
%
\font\numbers=cmss12
%\font\numbers=cmu10 scaled\magstep1
\font\upright=cmu10 scaled\magstep1
\def\stroke{\vrule height8pt width0.4pt depth-0.1pt}
\def\topfleck{\vrule height8pt width0.5pt depth-5.9pt}
\def\botfleck{\vrule height2pt width0.5pt depth0.1pt}
\def\Zmath{\vcenter{\hbox{\numbers\rlap{\rlap{Z}\kern
0.8pt\topfleck}\kern 2.2pt
                   \rlap Z\kern 6pt\botfleck\kern 1pt}}}
\def\Qmath{\vcenter{\hbox{\upright\rlap{\rlap{Q}\kern
                   3.8pt\stroke}\phantom{Q}}}}
\def\Nmath{\vcenter{\hbox{\upright\rlap{I}\kern 1.7pt N}}}
\def\Cmath{\vcenter{\hbox{\upright\rlap{\rlap{C}\kern
                   3.8pt\stroke}\phantom{C}}}}
\def\Rmath{\vcenter{\hbox{\upright\rlap{I}\kern 1.7pt R}}}
\def\Z{\ifmmode\Zmath\else$\Zmath$\fi}
\def\Q{\ifmmode\Qmath\else$\Qmath$\fi}
\def\N{\ifmmode\Nmath\else$\Nmath$\fi}
\def\C{\ifmmode\Cmath\else$\Cmath$\fi}
\def\R{\ifmmode\Rmath\else$\Rmath$\fi}
\def\Deln{\Delta_n}
%%%%%%%%%%%%%%%%%%%%%%%%%%%%%%%%%%%%%%%%%%%%%%%%%%%%%%%%%%%%%%%%%
 %%%%%%%%%%%%%%%%%% END OF DEFINITIONS %%%%%%%%%%%%%%%%%%%%%%
 %%%%%%%%%%%%%%%%%%%%%%%%%%%%%%%%%%%%%%%%%%%%%%%%%

%\section{Introduction}

\def\barray{\begin{eqnarray}}
\def\earray{\end{eqnarray}}
\def\beq{\begin{equation}}
\def\eeq{\end{equation}}

\def\gpar{g_\parallel}
\def\gperp{g_\perp}

\def\Jb{\bar{J}}
\def\dx{\frac{d^2 x}{2\pi}}

%*\section{Introduction}

The Renormalization Group (RG) continues to be one of the most important
tools for studying the qualitative and quantitative properties of quantum
field theories and many-body problems in Condensed Matter physics.
The emphasis so far has been mainly on flows toward fixed points in
the UV or IR.  Recently, an entirely novel kind of RG flow has been
discovered in a number of systems wherein the RG exhibits a cyclic
behavior:  after a {\it finite} RG transformation the couplings return to their
original values and the cycle repeats itself. 
Thus if one decreases the size of the system by a specific factor
that depends on the coupling constants, one recovers the initial system,
much like a Russian doll, or quantum version of the Mandelbrot set.
    Bedaque, Hammer
and Van Kolck observed this behavior in a 3-body hamiltonian
of interest in nuclear physics~\cite{nuclear}.  This motivated
Glazek and Wilson to define a very simple quantum-mechanical
hamiltonian with similar properties~\cite{Wilson}.
 In the meantime such behavior was proposed for a certain
regime of anisotropic current-current interactions
 in 2 dimensional quantum
field theory~\cite{BL}.

\def\dep{\delta}
\def\vep{\varepsilon}

The models in \cite{nuclear, Wilson} are problems in zero-dimensional
quantum mechanics, and are thus considerably simpler than the 
quantum field theory in \cite{BL}.  
 In the latter, standard quantum field theory methods of the
renormalization group were used,  however knowledge of the
beta function to all orders was necessary to observe the cyclic flow. 
What is somewhat surprising is that the model considered in \cite{BL} 
is not very exotic, and is in fact a well-known theory that
arises in many physical problems:  at one-loop it is nothing more
than the famous Kosterlitz-Thouless RG flow, where the cyclic
regime corresponds to $|\gperp| >  |\gpar|$. 
This motivated us to find a simpler many-body problem that
captures the essential features of the cyclic RG behavior.  
We found that a simple extension of the BCS hamiltonian has
the desirable properties.  Namely,  our model is based on the
BCS hamiltonian with scattering potential $V_{jj'}$ equal
to $ g+ i \theta$ for $\vep_j > \vep_{j'}$ and 
$g -i \theta$  for $\vep_j < \vep_{j'}$ in units of the energy
spacing $\dep$.

The main features of the spectrum are the following.  
For large system size,  there are an infinite number of 
BCS condensates, each characterized by an energy gap
$\Delta_n$ which depends on $g$, $\theta$.  The role of these many 
condensates becomes clearer when we investigate the RG properties. 
As in the models considered in \cite{nuclear,Wilson,BL}, the
RG flow possesses jumps from $g=+\infty$ to $g=-\infty$ and
a new cycle begins.   
Let $L = e^{-s} L_0$ denote the RG scale, which in our problem corresponds to 
$N$ the number of unperturbed energy levels, 
and $\lambda$ the period of an RG cycle:
$g(e^{-\lambda} L ) = g(L)$.  We show that $\lambda = \pi / \theta$.

The model we shall consider is an extension
of the reduced BCS model used to describe
ultrasmall superconducting grains \cite{vDR}, although
our results are valid for more general cases.
Let $c_{j,\pm}^\dagger$ 
($c_{j,\pm}$) denote creation-annihilation
operators for electrons in time reversal states $|\pm \rangle$.
The index $j=1,..,N$ 
refers to $N$ equally spaced energy levels $\vep_j$ with
$-\omega < \vep_j < \omega$.  
The energy $\vep_j$  represents the energy of 
a pair of electrons in
a given level.
The level spacing will be denoted 
$2 \dep$, i.e. $\vep_{j+1} - \vep_j = 2 \dep$, so that
$\omega = N \dep$ is twice the Debye energy. The energies
in this paper are twice their standard values \cite{largeN}. 
Let $b_j = c_{j,-} c_{j,+}$, 
 $b_j^\dagger  = c_{j,+}^\dagger  c_{j,-}^\dagger$
denote the usual Cooper-pair operators.    
The Hilbert space $\CH_N$ 
is spanned by the combination of empty and occupied states.
At half-filling the dimension of the Hilbert space 
is the combinatorial number $C^N_{N/2}$.

Our model is defined by the reduced BCS hamiltonian 
\beq
\label{2}
H =  \sum_{j=1}^N \vep_j ~ b^\dagger_j b_j ~ - ~ 
\sum_{j,j' = 1}^N ~ 
V_{jj'} \> b^\dagger_j b_{j'} ,
\eeq
where $V_{jj'}$ is the scattering potential. In the usual BCS model, 
$V_{jj'}$ is taken to be a constant. Here we
add an imaginary part which breaks time reversal, 
\beq 
\label{3}
\begin{array}{lcl}
V_{jj'} = \left\{ 
\begin{array}{ll}
G + i \Theta & \quad \mbox{if} \quad \vep_j > \vep_{j'} \\
G            & \quad \mbox{if} \quad \vep_j = \vep_{j'} \\
G - i \Theta & \quad \mbox{if} \quad \vep_j < \vep_{j'}
\end{array}, 
\right.
& &
\begin{array}{l}
G = g\delta \\
\Theta = \theta\delta
\end{array}.
\end{array}
\eeq
This hamiltonian is hermitian since $V^*_{jj'} = V_{j'j}$.  
We consider the positive dimensionless couplings $g$ and $\theta$.

%*\section{The many-body problem}
%*\subsection{Condensates and Gaps}

The BCS variational ansatz for this model is 
\beq
\label{14}
|\psi_{\rm BCS} \rangle = \prod_{j=1}^N \( u_j + v_j ~ b^\dagger_ j \) 
|0\rangle .
\eeq
The mean-field treatment yields the well-known equations
\barray
u_j^2 & = & \inv{2} \( 1+ \frac{\xi_j}{E_j} \), 
\qquad
v_j^2  \ = \ \inv{2} e^{2i\phi_j} \( 1- \frac{\xi_j}{E_j} \),
\nonumber \\
E_j & = & \sqrt{\xi_j^2 + \Delta_j^2 },
\qquad\quad
\xi_j \ = \ \vep_j - \mu - V_{jj}, 
\label{15}
\earray
where $\Delta_j$ and $\phi_j$  satisfy the gap equation: 
\beq
\label{16} 
\tilde{\Delta}_j = \sum_{j' \neq j} V_{jj'} 
\frac{\tilde{\Delta}_{j'} }{E_{j'}} ,
\qquad
\tilde{\Delta}_j \equiv \Delta_j e^{i \phi_j }.
\eeq
The chemical potential equation in our case is 
satisfied with $\mu =0$. 
In the thermodynamic limit, $N\to \infty$ and
$\dep \to 0$, with fixed $\omega = N \dep$, 
the sums over $\vep_j$ become the integrals 
$\int_{-\omega}^\omega d\vep / 2\dep $. The gap equation turns into
\beq
\label{17} 
\tilde{\Delta} (\vep ) = g \int_{-\omega}^\omega 
\frac{d\vep'}{2} \frac{\tilde{\Delta}( \vep' )}{E(\vep' ) } 
+ i \theta 
\[ \int_{-\omega}^\vep - \int_{\vep}^\omega \] 
\frac{d\vep'}{2} \frac{\tilde{\Delta}( \vep' )}{E(\vep' ) },
\eeq 
where 
$\tilde{\Delta} (\vep ) = \Delta (\ep ) e^{i \phi (\vep )}$.

Differentiating (\ref{17}) with respect to $\vep$ yields 
\beq
\label{18} 
\frac{d\phi}{d\vep} = \frac{\theta}{E(\vep )},
\eeq
and the condition that $\Delta (\vep ) = \Delta$ is independent of
$\vep$.  The solution to eq.~(\ref{18}) can be taken to be
\beq
\label{19}
\phi (\vep ) = \theta \sinh^{-1} \frac{\vep}{\Delta}.
\eeq
Using eq.~(\ref{19}) in the gap equation (\ref{17}) gives
\beq
\label{20}
1 =  \int_0^{\phi(\omega)} 
\frac{d\phi}{\theta} \( g \cos \phi + \theta \sin \phi \) 
\Longrightarrow 
\tan \phi(\omega) = \frac{\theta}{g}.
\eeq

Solving eq.~(\ref{20}) for the gap yields 
an infinite number of solutions $\Delta_n$.  They can
be parameterized as follows:
\beq 
\label{21}
\Delta_n = \frac{\omega}{\sinh t_n },
\quad 
t_n = t_0 + \frac{n\pi}{\theta}, 
\quad
n= 0,1,2,\ldots,
\eeq
where $t_0$ is the principal solution to the equation
\beq
\label{22} 
\tan ( \theta t_0 ) = \frac{\theta}{g},
\qquad
 0 < t_0 < \frac{\pi}{2\theta}. 
\eeq

The gaps  satisfy $\Delta_0 > \Delta_1 > \cdots$. 
Each gap $\Delta_n$ represents a different 
BCS eigenstate $|\psi_{\rm BCS}^{(n)} \rangle$. 
One can show that
$|\langle \psi_{\rm BCS}^{(n)} | \psi_{\rm BCS}^{(n')} \rangle | 
< \exp \left[ -N (\Delta_n - \Delta_{n'})^2/8 \omega^2 \right]$,
in the limit where $\Delta_n \ll \omega$.  Thus in the large~$N$ limit,
these eigenstates are orthogonal and should 
all appear in the spectrum, together
with the usual quasi-particle excitations above them. 
In the limit $\theta \rightarrow 0$ the gaps $\Delta_{n > 0} \rightarrow 0$,
and since $t_0 = 1/g$, $\Delta_0 \sim 2\omega e^{-1/g}$,
in the weak coupling regime, recovering the standard BCS result.

For weak coupling models $\Delta_n \ll \omega$, 
all the gaps are related by a scale transformation
$\Delta_n \sim 2 N \delta e^{-t_0 - n \pi/\theta}$.
Therefore, defining the condensation energy of the 
\mbox{n-th} BCS eigenstate as
$E_C^{(n)} = \langle \psi_{\rm BCS}^{(n)} | H | \psi_{\rm BCS}^{(n)}
 \rangle - E_{FS}$ we get
\beq
E_C^{(n)} \sim - \frac{\Delta_n^2}{8 \delta}
~
\Longrightarrow
~
E_C^{(n)} \sim - \frac{1}{2} \delta N^2 e^{-2t_0 - 2n\pi/\theta}.
\label{scaling}
\eeq
Thus  the spectrum of  condensation energies reflects  the 
scaling behavior of the gaps.

%*\subsection{Renormalization Group}

Next we derive RG equations for our model.  
Let $g_N$, $\theta_N$ denote the couplings for the hamiltonian
$H_N$ with $N$ energy levels.  The idea behind the RG method
is to derive an effective hamiltonian $H_{N-1}$ depending 
on renormalized couplings $g_{N-1}$, $\theta_{N-1}$ by integrating
out the highest energy levels $\vep_N$ or $\vep_1$.
This can be accomplished by a canonical transformation, 
which is formally analogous to the one used to 
derive the $t-J$ model from the Hubbard model
at strong coupling~\cite{hubbard}. 

We perform the calculation for general $V$.  
The integration of the level $\vep_N$ yields
\beq
\label{28}
V_{jj'}^{(N-1)} = V_{jj'}^{(N)} + \inv{2} 
V_{jN}^{(N)} V_{Nj'}^{(N)} \left( \inv{{\xi}_N - {\xi}_j} + 
\inv{{\xi}_N - {\xi}_{j'} } \right),
\eeq
where  ${\xi}_j = \vep_j - V_{jj}$. Integration 
of the level $\vep_1$ gives the same eq.~(\ref{28})
with the replacement 
${\xi}_N - {\xi}_j \rightarrow 
- {\xi}_1 + {\xi}_j$.

Specializing to the potential eq. (\ref{3}) and approximating
$\vep_N - \vep_j$ or  $- \vep_1 + \vep_j$ by $\omega = N \dep$, 
the above equation implies
\beq
\label{29} 
g_{N-1} = g_N + \inv{N} (g_N^2 + \theta_N^2 ),
\qquad
\theta_{N-1} = \theta_N.
\eeq
Thus $\theta$ is unrenormalized.  

In the large $N$ limit one
can define a variable $s= \log N_0/N$, where $N_0$ is the 
initial size of the system.  Then the beta function reads
\beq
\label{30} 
\frac{dg}{ds} =  (g^2 + \theta^2),
\qquad
s\equiv \log \frac{N_0}{N}.  
\eeq
The solution to the above equation is
\beq
\label{31}
g(s) = \theta \, \tan \left[ \theta s + \tan^{-1} 
       \left(\frac{g_0}{\theta} \right) \right], ~~ g_0 = g(N_0). 
\eeq

The main features of this RG flow  are the cyclicity
\beq
\label{32}
g(s + \lambda ) = g(s) ~ \Longleftrightarrow  ~ 
g(e^{-\lambda} N ) = g(N),
\quad 
\lambda \equiv \frac{\pi}{\theta},
\eeq
and the jumps from $+\infty$ to $-\infty$, when reducing the size.

The cyclicity of the RG has some important implications for the
spectrum.  Let $\{ E(g,\theta, N) \}$ denote the energy
spectrum of the hamiltonian $H_N$.  The RG analysis implies we can
compute this spectrum using the hamiltonian $H_{N'} (g(N') )$ if
$g(N') $ is related to $g(N)$ according the RG equation (\ref{31}). 
Moreover, if $N'$ and $N$ are related by one RG cycle,
$N' = e^{-\lambda } N$, then $g(N') = g(N)$.  Thus a plot of
the spectrum $\{ E(g,\theta, N) \}$ as a function of $N$ but
at {\it fixed} $g$, $\theta$ is expected to reveal the
cyclicity $\{ E(g,\theta,e^{-\lambda} N) \} = \{ E(g,\theta, N) \}$.
Since our RG procedure is not exact, we expect to observe
this signature within the range of our approximations, 
i.e.\ for $|E| \ll \omega$. Indeed, this agrees with the result shown in
eq.~(\ref{scaling}).
This can also be observed in fig.~1 for the one Cooper pair case, with 
the cyclicity given by $\lambda_1 = 2 \lambda$ (see below).
 
Eliminating $g_0$ in eq.~(\ref{31}) in terms of the mean-field 
solution, eqs.~(\ref{21},\ref{22}), 
we observe that the jumps in $g(s)$ 
from $+\infty$ to $-\infty$ occur at scales $s = t_n$.  
As $N$ decreases, $g$ increases steadily to $+\infty$ 
and then jumps to $-\infty$.  
At $g=+\infty$, $t_0 = 0$, $t_1 = \pi/\theta,\ldots$,
whereas for $g= -\infty $, $t_0 = \pi/\theta$, $t_1 = 2\pi/\theta,\ldots$.
Plugging this into eq.~(\ref{21}), one readily sees that 
\begin{eqnarray}
\Delta_0 (g = +\infty) & = & \infty  \nonumber \\
\Delta_{n+1} (g=+\infty ) & = &\Delta_n (g= -\infty ),
\label{33}
\end{eqnarray}
which indicates that at every jump the lowest condensate disappears 
from the spectrum, since $E_C^{(0)}(g = +\infty) = -\infty$.
Eq.~(\ref{33}) implies, for the remaining condensates, 
$E_C^{(n+1)}(g = +\infty) = E_C^{(n)}(g = -\infty)$. This result
is in agreement with eq.~(\ref{scaling}).
Therefore, the condensate $|\psi_{\rm BCS}^{(n+1)}\rangle $ of one
RG cycle plays the same role as  $|\psi_{\rm BCS}^{(n)}\rangle$ of
the next cycle.

The blow up of  $\Delta_0$ and 
$E_C^{(0)}$ at $g=+ \infty$
is an artifact of the RG scheme used here,
since we can only trust the RG for energies
below the cutoff $\omega$. However
the disappearance of bound states
is correctly described by this RG
(see the one Cooper pair problem
for a more detailed discussion).

When $N=\infty$ the infinite number of condensates are all expected
to appear in the spectrum.  However at finite $N$ this is not possible
since the Hilbert space is finite dimensional.  One can use the
RG to estimate the number of condensates $n_C$ in the spectrum
as a function of $N$. From the discussion above
a condensate disappears from the spectrum for each RG cycle.  
Thus $n_C$ should simply correspond to the number of cycles in $\log N$:
\beq
\label{34}
n_C \sim \frac{\theta}{\pi} ~ \log N .
\eeq

%*\section{Numerical work}
%*\section{The one-Cooper-pair problem}

%\medskip

%*\subsection{Bound states}

So far we have found a close relationship between the spectrum
of our extended BCS hamiltonian in the mean-field approximation 
and the RG flow of the coupling constants.  In order to
get a further confirmation of our results we should compute the
spectrum for a finite size system.  
However, it is very difficult
to reach intermediate sizes for this model numerically, since the
dimension of the Hilbert space grows as $2^N /N^{1/2}$.
Fortunately, to this end, the similarities between the many-body case
and the case of 
 one Copper pair, in the presence of the Fermi sea, are widely known.

For one Cooper pair in the presence of the Fermi sea,
consider an eigenstate of the form 
$|\psi \rangle = \sum_j \psi_j \> b^\dagger_j |0\rangle$. 
The Schrodinger equation reads
\beq
\label{4}
(\vep_j - E) \psi_j =  G \psi_j +
(G + i \Theta ) \sum_{l<j} \psi_l 
+ (G -i \Theta ) \sum_{l>j} \psi_l ,
\eeq
with $\vep_j \in (0,\omega )$, i.e. the Fermi sea is not accessible 
for the pair. 
Then, in the large~$N$ limit, the sums $\sum_j$ are replaced
by integrals $\int_0^\omega d\vep /2 \dep $, leading to
\beq
\label{6}
(\vep - E) \psi (\vep ) = g \int_0^\omega \frac{d\vep'}{2} ~ \psi (\vep' ) 
+ i \theta \[ \int_0^\vep - \int_\vep^\omega \] \frac{d\vep'}{2} 
\psi( \vep' ).
\eeq

Differentiating the above with respect to $\vep$ and integrating,
one obtains 
\beq
\label{7}
\psi (\vep ) \sim \frac{1}{\vep - E} e^{i \theta \log ( \vep - E)}.
\eeq
This wave function does not have cuts in two cases, 
i) $E < 0$ and ii) $E > \omega$. Case i) corresponds
to the usual Copper pair problem where one is looking
for bound state solutions (these are the solutions we
claim to have a similar behavior to the many-body case). 
Plugging eq.~(\ref{7}) back into
eq.~(\ref{6}) one finds 
\beq
\label{8}
\frac{g+ i\theta}{g-i \theta} = \( 1 - \frac{\omega}{E} \)^{i\theta}.
\eeq
This equation has an infinite number of solutions given by 
\beq
\label{9}
E_n = - \frac{ \omega}{e^{t_n} - 1 },
\qquad
t_n = t_0 + \frac{2 \pi n}{\theta}, \quad n \in {\cal Z},
\eeq
where $t_0$ is the principal solution to the equation
\beq
\tan \left(\frac{1}{2}\theta t_0 \right) = \frac{\theta}{g},
\qquad
0 < t_0 < \frac{\pi}{\theta}.
\label{11}
\eeq
The  $n \geq 0$ solutions correspond to $E_n < 0$, while those with 
$n < 0$ yield  $E_n > \omega$. 
%Again the $\theta \rightarrow 0$ reproduced the standard results.
%Again we recover the standard one-Cooper pair result 
%$E_{n \ne 0} \rightarrow 0$, 
%$E_0 \sim -\omega e^{-2/g}$ 
%in the $\theta \rightarrow 0$ limit.

As for the many-body case the spectrum has a scaling behavior 
for weak coupling systems, namely
\beq
E_n \sim - N \delta \; e^{ - t_0 - 2n\pi/\theta}.
\label{13}
\eeq 

An RG analysis similar to the one in the many-body problem
leads to the equation
\beq
\label{35} 
g_{N-2} = g_N + \frac{g_N^2 + \theta_N^2}{N-1-g_N},
\qquad
\theta_{N-1} = \theta_N,
\eeq
which in the large $N$ limit becomes
\beq
\frac{dg}{ds} = \inv{2} ( g^2 + \theta^2 ).
\label{36} 
\eeq
The solution to this equation is given by eq.~(\ref{31}) just by replacing 
$\theta s \rightarrow \theta s/2$.  This implies that
the period of the cyclicity in $s= \log N_0/N$ is
$\lambda_1 = 2\pi / \theta$.  

The factor $1/2$ in the above formulas, as compared
to the many-body problem, comes from the non accessibility
of the Cooper pair to the states below the Fermi level.

The discussion leading to eq.~(\ref{33}) can be repeated
for the one-Cooper pair problem where
the role of the n-th condensate is played by the n-th bound state
with energy $E_n$ given in eq. (\ref{9}), obtaining 
\begin{eqnarray}
E_0 (g = +\infty) & = & - \infty \nonumber \\
E_{n+1} (g= + \infty ) & = & E_n (g=-\infty ).
\label{37}
\end{eqnarray}

Thus we expect that in each RG 
cycle a bound state will disappear.
The analogue of equation (\ref{34}) is 
\beq
\label{38}
n_B \sim \frac{\theta}{2\pi} \, \log \frac{N}{2},
\eeq
where $n_B$ is the number of bound states in the spectrum. 

This again shows the agreement between the mean-field and 
the RG results.  We confirm bellow this picture with numerical calculations. 

\begin{figure}[ht!]
\begin{center}
\includegraphics[height= 5.4 cm,angle= 0]{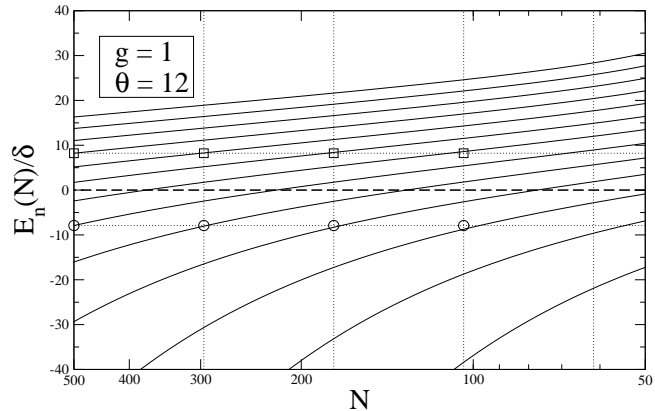}% Here is how to import EPS art
\end{center}
\caption{Exact eigenstates of one-Cooper pair Hamiltonian
for $N$ levels, from $N_0=500$ down to 50. We 
depict only the states nearest to zero. The vertical lines
are at the values $N_n=e^{-n \lambda_1} N_0$. The dotted 
horizontal lines show the cyclicity of the spectrum.   
}
\label{fig1}
\end{figure}

Fig.~1 shows the numerical solution of 
eq.~(\ref{4}) for $g=1$, $\theta =12$
and $N$ ranging from $500$ down to $50$. 
For each $N$
there are $n_B(N)$ bound states $E_n <0$,  
where $n_B(N)$ is in good agreement with eq.~(\ref{38}). 
 
The spectrum shows the self-similarity found in the approaches
above:  scaling the system  
by a factor $e^{- \lambda_1}$, with  
$\lambda_1 = 2 \pi/\theta$, one recovers the same spectrum 
for sufficiently small energies, i.e. 
\beq
 E_{n+1}( N, g, \theta) = 
E_{n}( e^{- \lambda_1} N, g, \theta).
\label{12}
\eeq

Fig.~1 also shows the existence of critical values
$N_{c,n}$, in the intervals 
$( e^{- n \lambda_1} N,e^{- (n+1) \lambda_1} N)$, 
where the bound state closest to the Fermi level 
disappears into the ``continuum''. This effect
leads to the reshuffling of bound states, 
$n+1 \rightarrow n$,  observed in eq.~(\ref{12}). 
The critical sizes are also related by 
scaling, i.e. $N_{c,n}/N_{c,n+1} = e^{- \lambda_1}$.
All these phenomena are  in good agreement with
the RG interpretation we proposed,
where the condensates disappear at scales where
$g(s=t_n) = +\infty$.  All these points are related by the
scaling factor $e^{-\lambda_1}$.

The RG behavior is presented in
fig.~2, which shows the eigenvalues $E_n(N)$ 
of the one-Cooper pair Hamiltonian, with 
$g_N$ running under eq.~(\ref{35}). The spectrum 
remains unchanged for $E_n(N) \ll N \delta $, as shown in fig.~2b. 
In fig.~2a one observes that  
for the energies $E_n(N) \gtrsim N \delta $ 
the result of the RG is not reliable. 
Nevertheless the RG flow describes 
qualitatively the disappearance of the lowest
bound state and the reshuffling of energy levels 
after a cycle, and furthermore at the predicted scales.

%\medskip

In summary, we have shown that
adding to the standard BCS Hamiltonian
a time reversal breaking term, 
parametrized by a coupling constant $\theta$,
generates an infinite number of
condensates with energy gaps $\Delta_n$ 
related, for weak BCS couplings $g$,  
by a scale factor  $e^{- \lambda}$
with $\lambda = \pi/\theta$. This 
unusual spectrum is explained by  
the cyclic behavior of the
RG flow of  $g$, which
reproduces itself after a finite RG time $s$ equal to  $\lambda$. 
We have also solved the finite temperature BCS gap
equation, obtaining a critical temperature 
$T_{c,n}$ for the n-th condensate which is related
to the zero temperature gap $\Delta_n(0)$ exactly
as in the BCS theory, i.e.   
$\Delta_n(0)/T_{c,n} \cong 3.52$ for weak couplings.  

The simplicity of the model proposed in this letter
suggests that {\em Russian doll superconductors} could
perhaps be realized experimentally. Finally, we point out that
the critical temperature can be raised by varying $\theta$.

\begin{figure}[h]
\begin{center}
\includegraphics[height= 5.5 cm,angle= 0]{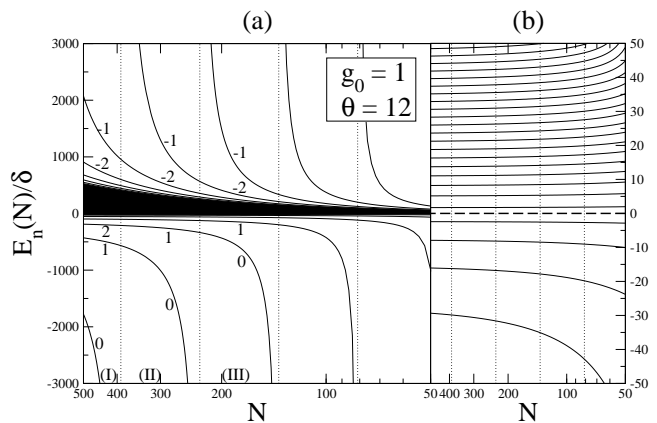}% Here is how to import EPS art
\end{center}
\caption{Eigenstates of one-Cooper pair Hamiltonian 
with $g_N$ given by eq.(\ref{35}) with $g_{0}=1$ and $\theta=12$.
The vertical lines denote the positions at which  
$g$ jumps from $+\infty$ to $- \infty$. 
} 
\label{fig2}
\end{figure}

We thank D. Bernard, J. Dukelsky and M.A. Mart\'{\i}n-Delgado 
for discussions. 
This work has been supported by the Spanish grants 
SAB2001-0011 (AL), 
BFM2000-1320-C02-01 (JMR and GS), and by the NSF of the USA.

\end{document}